\documentclass[12pt,a4paper,oneside,headings=small]{scrartcl}
\usepackage[utf8]{inputenc}
\usepackage[english]{babel}
\usepackage{graphicx,geometry}
\usepackage{fancyhdr,mathptmx,float}
\usepackage[T1]{fontenc}

\usepackage{helvet,array}
\newcolumntype{L}[1]{>{\raggedright\let\newline\\\arraybackslash\hspace{0pt}}m{#1}}
\usepackage[x11names]{xcolor}
\usepackage{booktabs}
\usepackage{parskip}
\usepackage{subfig}
\usepackage{amsmath}
\usepackage{bm}
\usepackage{url}
\usepackage{upgreek}
\usepackage[comma,authoryear]{natbib}
\usepackage{setspace}
\usepackage{tikz}
\usetikzlibrary{shapes,shadows,arrows,positioning,patterns,decorations.pathreplacing,calc}

\fancypagestyle{firstPage}{%
\fancyhead[L]{}
\fancyfoot[L]{{\textit{\fontsize{9}{7}\selectfont 15$^{th}$ International Symposium on Hazards, Prevention and Mitigation of Industrial Explosions\\
Naples, ITALY -- June 10-14, 2024}}}
\fancyfoot[R]{\raisebox{-7mm}{\includegraphics[trim=0mm 0mm 0mm 0mm,clip=true,width=0.20\textwidth]{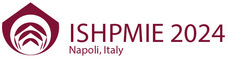}}}

}

\setkomafont{section}{\normalfont\bfseries}
\RedeclareSectionCommand[
  beforeskip=-1\baselineskip,
  afterskip=.1\baselineskip]{section}
\setkomafont{subsection}{\normalfont\itshape}
\RedeclareSectionCommand[
  beforeskip=-1\baselineskip,
  afterskip=.1\baselineskip]{section}
\setkomafont{subsubsection}{\normalfont\itshape}
\RedeclareSectionCommand[
  beforeskip=-1\baselineskip,
  afterskip=.1\baselineskip]{section}

\addto\captionsenglish{
        
        }
\usepackage[format=plain,
            labelfont=it,
            textfont=it]{caption}

\begin{document}
\thispagestyle{firstPage}

\begin{center}
{\setstretch{1.8}
\textbf{\LARGE Unmasking hidden ignition sources: A new approach to finding extreme charge peaks in powder processing}\\[15pt]
}
Holger Grosshans$^{a,b}$, Wenchao Xu$^a$, Simon Janta\v{c}$^a$ \& Gizem Ozler$^{a,b}$\\[5pt]
$^a$ Physikalisch-Technische Bundesanstalt (PTB), Braunschweig, Germany\\
$^b$ Otto von Guericke University of Magdeburg, Institute of Apparatus and Environmental Technology, Magdeburg, Germany\\[5pt]
E-mail: {\color{blue} \textit{holger.grosshans@ptb.de}}\\[5pt]
\end{center}

\medskip
\bigskip
\section*{Abstract}

Powders acquire a high electrostatic charge during transport and processing.
Consequently, in the aftermath of dust explosions, electrostatic discharge is often suspected to be the ignition source.
However, definite proof is usually lacking since the rise of electrostatic charge cannot be seen or smelled, and the explosion destroys valuable evidence.
Moreover, conventional methods to measure the bulk charge of powder flows, such as the Faraday pail, provide only the aggregate charge for the entire particle ensemble.
Our simulations show that, depending on the flow conditions, contacts between particles lead to bipolar charging.
Bipolar charged powder remains overall neutral;
thus, a Faraday pail detects no danger, even though individual particles are highly charged.
To address this gap, we have developed a measurement technology to resolve the powder charge spatially.
The first measurements have revealed a critical discovery: 
a localized charge peak near the inner wall of the conveying duct is 85 times higher than the average charge that would be measured using the Faraday pail.
This finding underscores the possibility of extremely high local charges that can serve as ignition sources, even though they remain undetected by conventional measurement systems. 
Our new technology offers a solution by spatially resolving the charge distribution within powder flows, unmasking hidden ignition sources, and preventing catastrophic incidents in the industry. 

\medskip
\noindent
\textbf{Keywords:} \textit{industrial explosions, powder processing, electrostatics, measurement, simulation}

\section{Introduction}

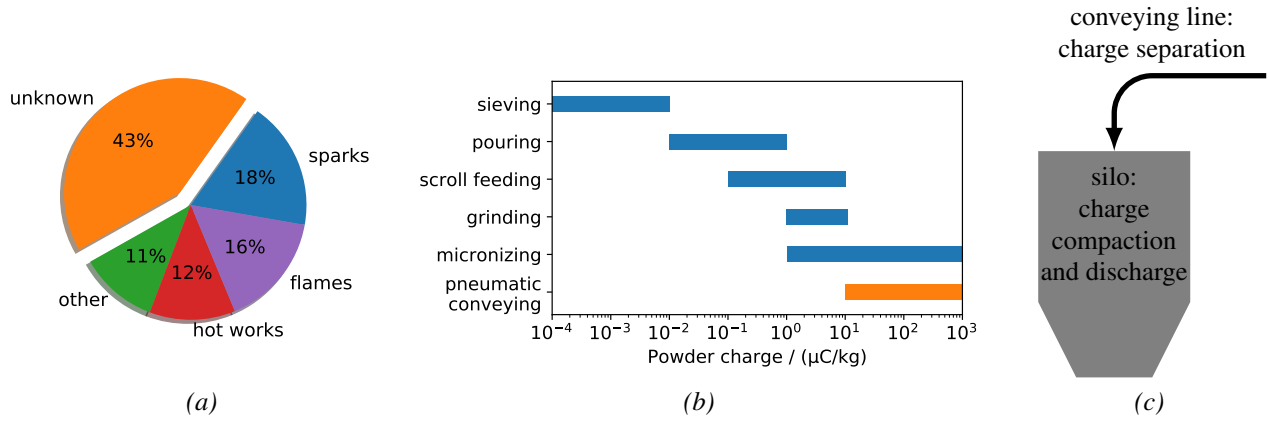
\begin{figure}[tb]
\begin{center}
\subfloat[]{\includegraphics[trim=5mm 10mm 0mm 0mm,clip=true,width=0.32\textwidth]{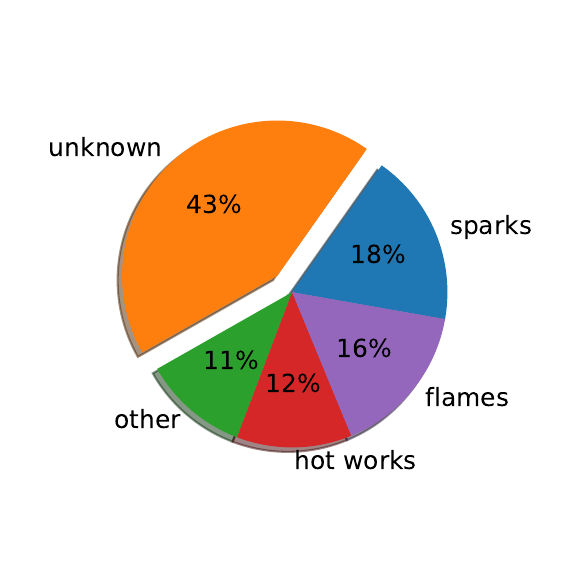}\label{fig:ignition-source}}
\subfloat[]{\includegraphics[trim=0mm 0mm 0mm 0mm,clip=true,width=0.45\textwidth]{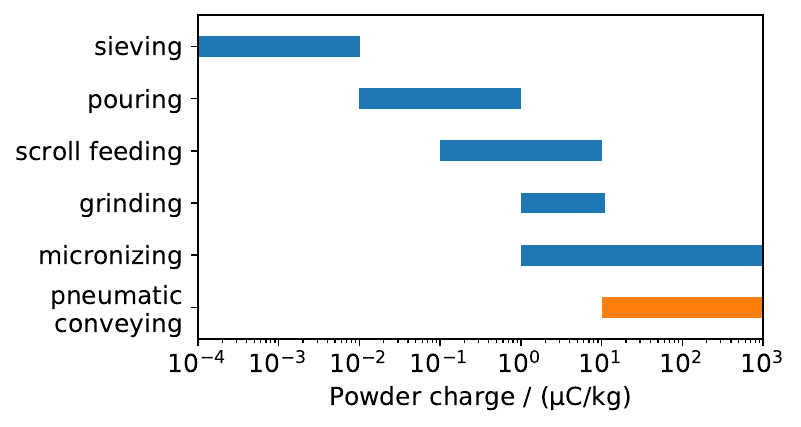}\label{fig:operation}}
\quad
\subfloat[]{
\begin{tikzpicture}
\draw [->,>=latex,line width=2] (3,1) -- (1.5,1) node [above,align=center,font=\footnotesize] {conveying line:\\charge separation} to [out=180,in=90] (1,.5) -- (1,0);
\path [fill=gray] (0,0) -- (2,0) -- (2,-2) -- (1.5,-3) -- (0.5,-3) -- (0,-2);
\node [align=center,font=\footnotesize] at (1,-1) {silo:\\charge\\compaction\\and discharge};
\end{tikzpicture}
\label{fig:compaction}
}
\end{center}
\caption{
(a) 
The ignition source of 43\% of dust explosions around the world from 1785 to 2012 (data collected by~\citet{Zhi15} and compiled by \citet{Newson21}) is unknown.
(b) pneumatic conveying generates high electrostatic charge, which then (c) compacts and often discharges in silos.}
\label{fig:intro}
\end{figure}

%1- 40\% of all dust explosions ignition source unknown
For many dust explosions \citep[43\% according to][]{Zhi15,Newson21}, the ignition source remains unclear (figure~\ref{fig:ignition-source}).
When investigators fail to identify the source of an explosion, they often suspect electrostatic discharges because electrostatic discharges leave little or no traces.
They cannot be smelled or felt, so they are not reported by eyewitnesses.
Further, as we will elaborate in this article, conventional measurement equipment cannot detect the local rise of charge.
Thus, evidence is usually lacking.

%2- pneumatic conveying leads to highest powder charge
% where does the actual spark and igniion happen? in the pipe? brush and cone discharge, particle relaxation
In the pursuit of finding processes or locations in industrial plants where charge separates at high rates, pneumatic conveyors are a top candidate.
Pneumatic conveying, as other powder operations, induces tribocharging through frictional contact of particles with pipes and ducts~\citep{Gro23a}.
The conveying pipeline, bends, and other components act as charge generators, resulting in the accumulation of electrostatic charges on the conveyed powder and flown-through devices.
Compared to other powder process operations, pneumatic conveying reaches several orders of magnitude higher charge levels~(figure~\ref{fig:operation}).

Being the operation that generates the most charge does not mean that pneumatic conveyors are at the highest risk of experiencing an explosion.
Instead, the charge generated in conveyors can cause discharges in the conveyor itself and any other downstream process.
The conveying pipeline itself, along with bends and transitions, is prone to discharges, but explosions are unlikely in this location if the equipment is electrically bonded and the dust concentration is low.
Even during close proximity between particles or between particles and the pipeline wall, electrostatic discharges occur due to differences in charge potential \citep{Mat95c}, but the energy of these discharges is small compared to the dust's Minimum Ignition Energy~(MIE).

%3- conveying line to silos
The potential dust explosion locations are downstream of pneumatic conveyors, where particles pile up or form layers, increasing the particle concentration and energy density.
Especially inside silos, combustible dust clouds can form, and cone discharges can lead to dust explosions \citep{Choi18}.
Silos are filled with products by trucks, bucket elevators, or pneumatic conveyors, all of which have the likelihood of producing a dust cloud during the filling process.
Particles that received a high charge in the conveying line can experience cone discharges due to charge compaction \citep{Glor84}.
Charge compaction describes the mechanical and electrostatic charge compression when filling a conveyed product into a silo.
If the powder's surface conductivity is low, the charge dissipates slowly toward the ground, and the charge density at the silo's bed becomes much higher than in the conveying system or the charged dust cloud \citep{Glor01}.
In other words, pneumatic conveying generates a high powder charge, and filling a silo compacts the charge to high densities (figure \ref{fig:compaction}).

%4- accident cases of dust explosions where electrostatics could have been the cause
%  best cases:
%  a. pneumatic conveyor, measurement showed low charge, still explosion
%  b. pneumatic conveyor present, ignition source fully unclear
%    - examples of silo explosions with conveying line, ignition source unclear
%  c. pneumatic conveyor gave a surprisingly high charge

Dust explosions in silos are not uncommon. 
In 2022, one-quarter of all industrial dust explosions happened in silos and other storages \citep{Clo23}, and the root of these explosions often remains unclear.
For example, the 2020 grain dust explosion at Tilbury docs in Essex, the largest grain terminal in the UK, destroyed several silos and produced 75-meter-high flames.
The fire brigade worked at the explosion site for 20 days, repeatedly extinguishing smoldering fires from the destroyed silos.
However, the source of the explosion was not found.
In the same year, the explosion at Ag Partners in Royal (Iowa, USA) destroyed a silo where a mixture of corn and soybean was stored along with the adjacent installations.
Again, the cause of the explosion is unknown.%refxx

%5- conventional techniques measure only average charge
While the roots for these dust explosions remain unknown, we do know that the measurement of electrostatic charge that is introduced from the conveying line to the storage has limitations.
Conventional Faraday pails, or cups, can be strategically placed at selected points in the pneumatic conveying system to detect the charge buildup during various conveying stages.
Faraday pails measure the charge induced on their inner surface; 
thus, they provide the total sum of the enclosed powder's electrostatic charge.
They cannot provide any other detailed quantity.
In other words, the charge read by a Faraday pail is the integral of the charge in the enclosed volume, $Q(t)=\int_\mathcal{V} Q({\bm x},t)$.
If part of the powder carries a positive and the other part a negative charge, the Faraday pail may return a close-to-zero reading.
Similarly, if the powder's charge at one duct wall is extremely high but low within the rest of the duct's area, the Faraday pail returns a mediocre overall charge.

The guidelines (\citet{TRBS,60079-32-1}) recommend measuring the electric field strength above the powder heap within the silo to assess the probability of cone discharges.
However, according to \citet{Glor13}, such measurement is too sophisticated for regular industrial practice.

We found, through simulations, that pneumatic conveying leads to highly distributed charge.
That means the powder can charge bipolar despite remaining overall neutral \citep{Gro23c}.
And, charge distributes in space, meaning that some flow regions become highly charged while others remain uncharged \citep{Gro17a}.
Experimental studies indicated that the spatial charge profiles generated during pneumatic conveying affect downstream processes.
More specifically, depending on whether the conveying line filled a silo from a central or a tangential location, the most frequently measured cone discharge nearly tripled from 50~nC to 140~nC \citep{Glor97}.
This increase might result from a differently distributed powder charge in the silo, generated during pneumatic conveying and propagated to the silo during filling.

As discussed above, today's measurement equipment cannot detect these charge distributions.
Therefore, we developed a new measurement technology that can resolve so-far hidden charge peaks.
This paper reports on the simulations predicting distributed flow charge and the new measurement technology that can resolve these distributions.

\section{Simulations predicting bipolar and spatially distributed charge}

Our simulations found powder transported by turbulent air to charge nonuniformly.
The distributed charge presented in the following cannot be detected by today's measurement equipment, which provides only the integral charge.
The mathematical model and numerical methods of our simulation tool \citet{pafiX} are described in detail by \citet{Gro20d} and are summarized in the following.

\subsection{Mathematical model and numerical methods}

\citet{pafiX} employs a four-way coupled Eulerian-Lagrangian approach that tracks each particle individually.
The particle motion is sensitive to perturbations;
therefore, turbulence is modeled by DNS.
Further, the interaction between charged particles and the incident electric field is captured by advanced methods as outlined below.

% fluid
The Navier-Stokes equations for constant densities and diffusivities,
\begin{subequations}
\begin{equation}
\label{eq:mass}
\nabla \cdot {\bm u}_{\textrm g} \;=\;0\,,
\end{equation}
\begin{equation}
\label{eq:mom}
\frac{\partial {\bm u}_{\textrm g}}{\partial t} + ({\bm u}_{\textrm g} \cdot \nabla) {\bm u}_{\textrm g}
\;=\; - \frac{1}{\rho} \nabla p  + \nu \nabla^2 {\bm u}_{\textrm g} + {\bm f}_{\mathrm s} + {\bm f}_{\mathrm f} \,,
\end{equation}
\end{subequations}
describe the carrier gas flow.
Therein, ${\bm u}_{\textrm g}$ is the fluid velocity, $\rho$ the density, $p$ the dynamic pressure, and $\nu$ the kinematic viscosity.
Further, ${\bm f}_{\textrm s}$ is the source term for the momentum transfer from the particles to the gas.
To simulate pneumatic conveying, we model a short section of the duct, apply periodic boundary conditions in streamwise ($x$) direction, and drive the flow by the pressure gradient, ${\bm f}_{\textrm f}$.
In the above equations, second-order schemes approximate the spatial and temporal derivatives.

% particles
For each particle of the mass $m$, we solve the acceleration, ${\bm a}$, based on Newton's second law of motion in the Lagrangian framework,
\begin{equation}
\label{eq:newton}
m {\bm a} \;=\; {\bm F}_{\mathrm{d}} + {\bm F}_{\mathrm{c}} + {\bm F}_{\mathrm{e}} + {\bm F}_{\mathrm{g}} \, .
\end{equation}
The drag force acting on the particle, ${\bm F}_{\mathrm{d}}$, is computed according to the empirical correlation provided by \citet{Put61}.
The collisional force, ${\bm F}_{\mathrm{c}}$, includes particle-particle and wall-particle collisions.
To detect particle-particle collisions, we implemented ray casting \citep{Roth82,Schr01}, a variant of the hard-sphere approach.

A hybrid method \citep{Gro17e} computes the electrostatic forces, ${\bm F}_{\mathrm{e}}$, on a particle carrying the charge $Q$.
This method adds the the Coulomb interactions between the particle and its neighboring particles to the far-field forces calculated with Gauss law.
The last term in equation~(\ref{eq:newton}), ${\bm F}_{\mathrm{g}}$, denotes the gravitational force.

\subsection{Simulation results}

\begin{figure}[b]
\begin{center}
\subfloat[$St=$ 20]{
\includegraphics[trim=0cm 0cm 0cm 0cm,clip=true,width=0.46\textwidth]{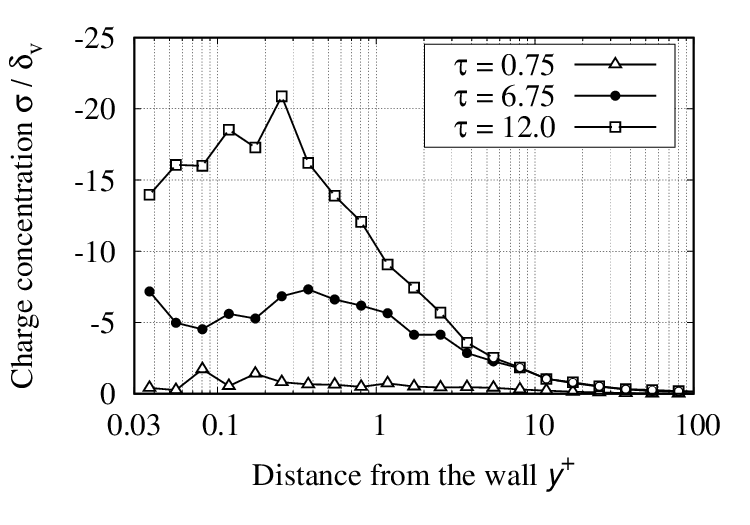}
\label{fig:pwa}
}
\quad
\subfloat[$St=$ 1.17, 2.34, 4.69]{
\includegraphics[trim=0cm 0cm 0cm 0cm,clip=true,width=0.48\textwidth]{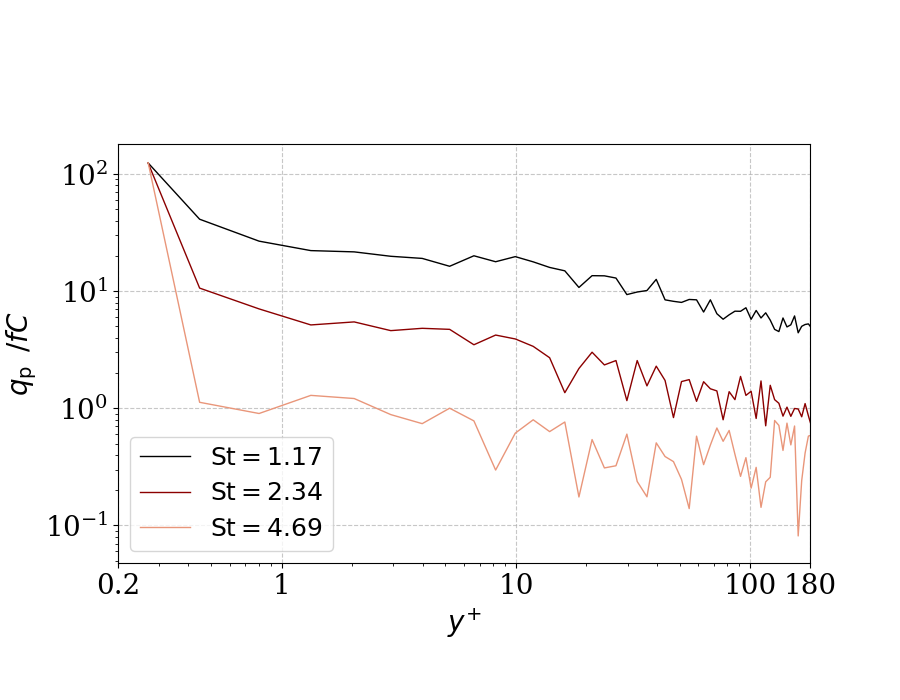}
\label{fig:pwb}
}
\end{center}
\caption{
Simulation of powder charge profiles during pneumatic conveying.
The extremes of the profiles close to the conveying duct's walls are hidden from conventional Faraday pails that provide the integral charge
\citep[(a) reprinted with permission from][]{Gro17a}.
}
\label{fig:pw}
\end{figure}

Figure~\ref{fig:pw} shows the spatial powder charge profiles from channel flow simulations.
The gas flow's frictional Reynolds number, $Re_\mathrm{\tau}=360$, is at the lower end but still representative of pneumatic conveying.

Figure~\ref{fig:pwa} depicts the temporal charge evolution of particles of a Stokes number of 20 ($St= \tau/\tau^+_\mathrm{f}$, where $\tau$ is the particle's reponse time and $\tau^+_\mathrm{f}$ the gas flow's frictional time scale).
For this simulation, the condenser model~\citep{Soo71,Mas76,John80} predicted the charge exchange between particles and walls and other particles.

Figure~\ref{fig:pwb} displays the charge profile of three different Stokes numbers, each when the powder reaches half its saturation charge.
Here, a simple, generic model computes the transfer of charge from a wall to a particle, $\Delta q_{\mathrm{pw},n}$, and between particles, $\Delta q_{\mathrm{pp},n}$,
\begin{equation}
\Delta q_{\mathrm{pw},n} = C_1 A_\mathrm{pw} \left( q_\mathrm{eq} - q_n \right)/A_\mathrm{p}
\quad
\text{and}
\quad
\Delta q_{\mathrm{pp},n} = C_2 A_\mathrm{pp} \left( q_m - q_n \right)/A_\mathrm{p} \, .
\end{equation}
In these equations, $q_n$ and $q_m$ denote the charge of the particles before the contact, $q_\mathrm{eq}$ their saturation charge, $A_\mathrm{pw}$ and $A_\mathrm{pp}$ the contact areas calculated by a Hertzian model, $A_\mathrm{p}$ the particle's total surface area, and $C_1$ and $C_2$ are constants.

In these simulations, no matter which charge model and Stokes number, the charge at the walls rises quickly while the charge in the bulk of the flow remains low.
In other words, the charge of the flow is spatially non-uniform regardless of the particle type or precise physical and chemical charging mechanisms.

As discussed in the introduction, a Faraday pail or induction probe cannot resolve the charge in space.
Thus, these charge peaks, which may be a potential ignition source, have been hidden from measurement.

Besides missing ignition sources, measuring only the integral of the charge comes with another severe disadvantage:
simulations lack validation data.
Simulation codes of powder charging have evolved enormously in recent years.
Now, they can resolve all turbulence scales and the dynamics of each particle.
Whereas experimentally derived, time-averaged, and spatially-resolved profiles of turbulence statistics and particle concentrations are standard to validate the simulations of uncharged particles, the equivalent data for powder charging is lacking.
The lack of time-averaged and spatially-resolved powder charge profiles hinders the validation and further development of simulation codes for powder flow charging.

\begin{figure}[tb]
\begin{center}
\subfloat[]{\includegraphics[width=0.47\textwidth]{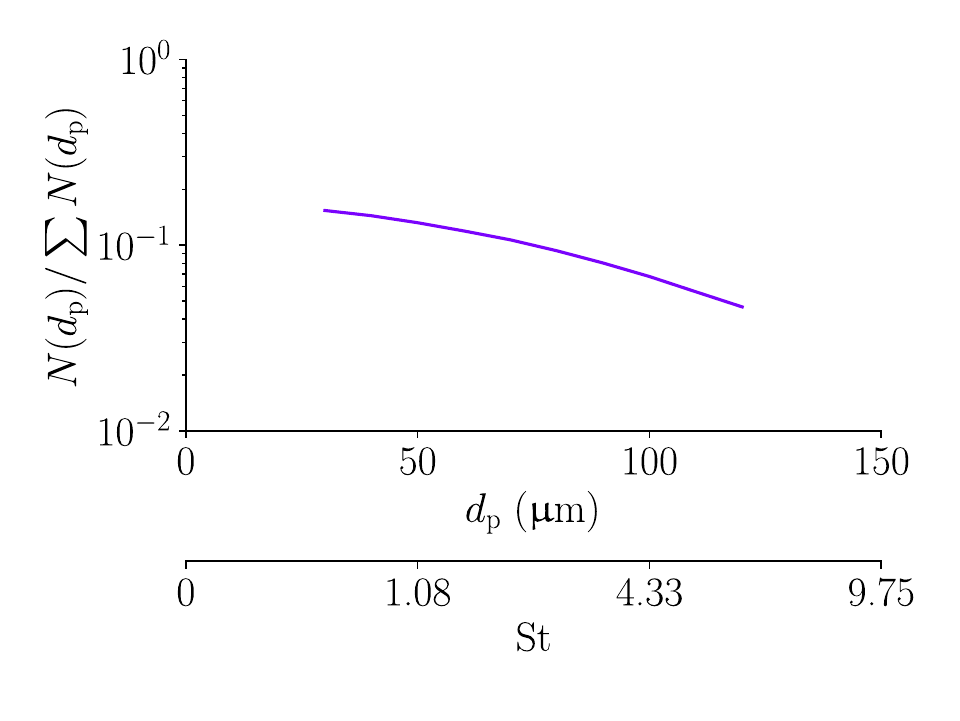}
\label{fig:ppa}
}
\quad
\subfloat[]{\includegraphics[width=0.47\textwidth]{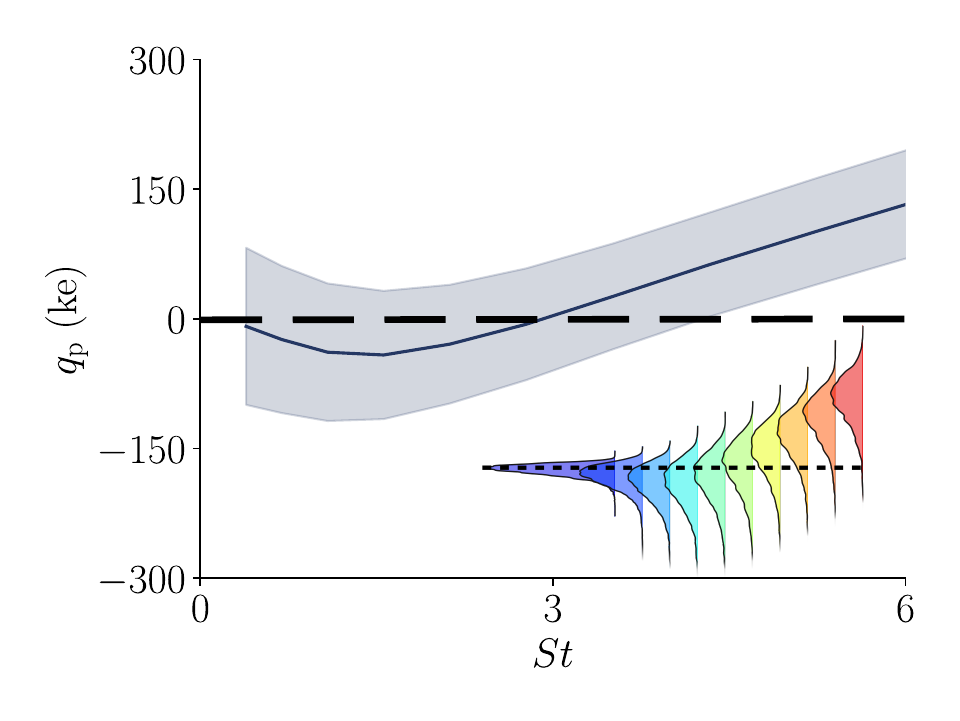}
\label{fig:ppb}
}
\end{center}
\caption{Simulation of bipolar powder charging during pneumatic conveying.
(a) Size distribution of polydisperse powder.
(b) The extremes of the particle charge distribution remain hidden from the averages provided by conventional Faraday pails~\citep{Gro23c}.
The inset gives the charge distribution per particle size.
}
\label{fig:pp}
\end{figure}

The simulation presented in figure~\ref{fig:pp} depicts bipolar powder charging.
Powder charges bipolar if its size distribution is polydisperse and particles of different sizes collide~\citep{Wait14}.

We computed the collisional charge exchange by the surface-state/mosaic model in the formulation of \citet{Kon17}.
According to this model, the charge
\begin{equation}
\Delta q_{\mathrm{pp},n} = \epsilon (c_{\mathrm{s},m}-c_{\mathrm{s},n})A_\mathrm{pp}
\end{equation}
transfers from particle $m$ to $n$ during collision.
In the above equation, $\epsilon$ denotes the charge of an electron or anion with a charge number of -1, and $c_{\mathrm{s},n}$ and $c_{\mathrm{s},m}$ the density of transferrable elementary charges available on the particles' surfaces.

The simulation of the particle size distribution in figure~\ref{fig:ppa} results in the charge distribution in figure~\ref{fig:ppb} \citep{Gro23c}.
We found that the turbulent carrier flow has a leading influence on the shape of the resulting charge distribution.
More specifically, mid-sized particles collect the most negative and large particles most positive charge.

Despite the high charge of both polarities collected by specific particle size classes, the charge of the overall powder remains neutral.
That means the integral charge read by a Faraday pail would be zero.
Therefore, the powder charge and the related hazard remain unnoticed.

\section{Experiments resolving spatially distributed charge}

\subsection{Test-rig \& measurement system}
\label{sec:rig}

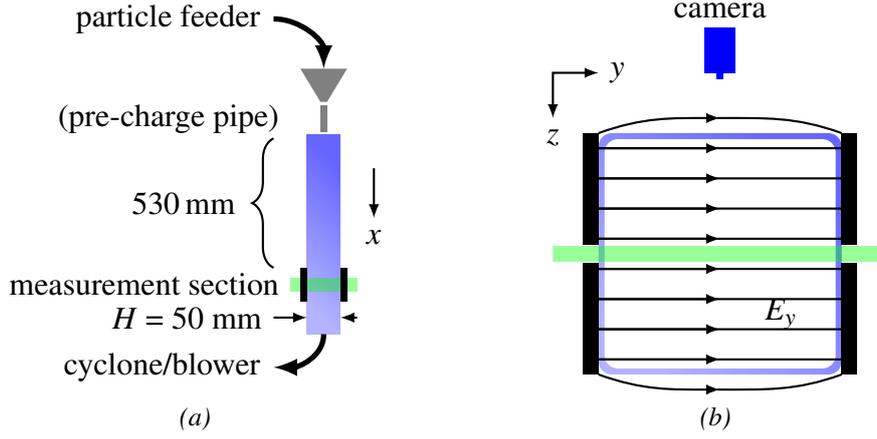
\begin{figure*}[tb]
\centering
\subfloat[]{
\begin{tikzpicture}[scale=0.44]
\draw [->,>=latex,line width=2] (-1,8.5) node [left] {particle feeder} to [out=0,in=90] (0.5,7);
\path [fill=gray] (.6,6) -- (1.2,7) -- (-0.2,7)-- (.4,6);
\path [fill=gray] (.4,5.9) rectangle (.6,5.1);
\node [left] at (-.5,5.5) {(pre-charge pipe)};
\path [shading = axis,rectangle, left color=blue!30!white, right color=blue!60!white,shading angle=135] (0,-1) rectangle (1,5);
\node [left] at (-.5,.5) {measurement section};
\path [fill=green,opacity=.4] (-.5,0.3) rectangle (1.5,0.7);
\path [fill=black] (0,0) rectangle (-.2,1);
\path [fill=black] (1,0) rectangle (1.2,1);
\draw [->,>=latex,line width=2] (.5,-1) to [out=270,in=0] (-1,-2) node [left] {cyclone/blower};
\draw [thick,decorate,decoration={brace,amplitude=8pt}] (-1,1) -- (-1,4.9) node[midway,left,xshift=-10]{530\,mm};
\draw [->,>=latex,thick] (2,4) -- (2,2.5) node [below] {$x$};
\draw [->,>=latex,thick] (-1,-.5)  node [left]{$H$ = 50 mm} -- (0,-.5);
\draw [<-,>=latex,thick] (1,-.5) -- (1.5,-.5);
\end{tikzpicture}
\label{fig:poexpa}
}
\qquad
\qquad
\subfloat[]{
\begin{tikzpicture}[scale=0.4]
\path [rounded corners=5pt,shading = axis,rectangle, left color=blue!30!white, right color=blue!60!white,shading angle=135] (0,0) rectangle (8,8);
\path [rounded corners=5pt,fill=white] (0.2,0.2) rectangle (7.8,7.8);
\draw [->,>=latex,thick] (0,0) to [out=-20,in=180] (4,-.5); \draw [,thick] (3.9,-.5) to [out=0,in=200] (8,0);
\draw [->,>=latex,thick] (0,8) to [out=20,in=180] (4,8.5); \draw [,thick] (3.9,8.5) to [out=0,in=160] (8,8);
\foreach \i in {0.5,...,7.5} \draw [->,>=latex,thick] (0,\i) -- (4,\i);
\foreach \i in {0.5,...,7.5} \draw [,thick] (0,\i) -- (8,\i);
\draw [green,line width=6,opacity=.4] (-1.5,4) -- (9.5,4);
\path [fill=blue] (3.5,10) rectangle (4.5,11.5) node [midway,above,yshift=8] {camera};
\path [fill=blue] (3.9,9.8) rectangle (4.1,10);
\path [fill=black] (-.5,0) -- (0,0) -- (0,3.7) -- (-.5,3.7);
\path [fill=black] (-.5,8) -- (0,8) -- (0,4.3) -- (-.5,4.3);
\path [fill=black] (8,0) -- (8.5,0) -- (8.5,3.7) -- (8,3.7);
\path [fill=black] (8,8) -- (8.5,8) -- (8.5,4.3) -- (8,4.3);
\node at (6,2) {$E_y$};
\draw [<->,>=latex,thick] (0,10) node [right] {$y$} -- (-1.5,10) -- (-1.5,8.5) node [below] {$z$};
\end{tikzpicture}
\label{fig:poexpb}
}
\caption[]{
(a) Pneumatic conveying pilot plant to test the new measurement system.
(b) Components of the measurement section: electric field ($E_y$), laser sheet (green), and PTV camera
\citep{Gro23b}.
}
\label{fig:poexp}
\end{figure*}

In response to the need for the resolved powder charge transported by a turbulent flow, we developed a new measurement technology and installed it at a laboratory-scale pneumatic conveyor (figure.~\ref{fig:poexpa}).
The measurement system's details and accuracy are discussed by \citet{Gro23b}.
In the following, we describe modifications to the system and further tests.

At the top of the conveyor, the feeder supplies particles to the duct's inlet.
Optionally, a narrow aluminum pipe assigns pre-charge to the particles before entering the duct.
According to our previous tests, the net pre-charge is positive~\citep{Gro23a}.
The duct conveys the powder vertically downward, aligned with the gravitational acceleration.
After the duct's outlet, a cyclone followed by a filter separates the particles from the airflow.
A blower sucks the air and powder through the rig.

The duct is made of transparent PMMA, providing optical access for Particle Tracking Velocimetry (PTV) (figure~\ref{fig:poexpb}).
The duct's length, from the inlet to the beginning of the electric field, is 530~mm.
When flowing downstream, the particles reach the measurement section, where two parallel plates generate an electric field ($E_y$) that can be switched on or off.
The plates are electrically conductive but transparent so the laser can pass through and illuminate the flow.
The duct has a square cross-section.
Each planar wall's inner side length is $H=$~50~mm.
The laser beam points in spanwise ($y$) direction.
In streamwise direction ($x$), the sheet begins 80~mm after the edge of the electric field and spans about 50~mm.

The tests were conducted at a frictional Reynolds number of 360, equal to the abovementioned simulations.
The particles were monodisperse, spherical, of a size of $d=100~\upmu$m, and made of PMMA.
Their mass flow rate was 0.2~g/s.
During the tests, the relative humidity and temperature in the lab were 52.5\% and 21.5~°C.

The measurement system derives the particles' charge from the force balance and their response to the electric field.
In $y$ direction, the longitudinal axis of the laser beam, the force balance of equation~(\ref{eq:newton}) reduces to
\begin{equation}
\label{eq:ae}
m a^E_y = F_{\mathrm{d},y} + F_{\mathrm{c},y} + F_{\mathrm{e},y} \, . 
\end{equation}
Here, the superscript $E$ indicates the influence of the electric field.
Gravitational forces vanish since the laser points horizontally.

The problem in solving equation~(\ref{eq:ae}) lies in the unknown terms $F_{\mathrm{c},y}$ and $F_{\mathrm{d},y}$.
The collisional force is generally unknown, but it can be locally large, even in dilute flows.
For particles of $St\ll\infty$, the drag force requires knowledge of the instantaneous gas velocity.
However, in turbulent flows, due to the chaotic fluctuations, the instantaneous gas velocity is unknown. 

We proposed a solution to this problem based on averaging,
\begin{equation}
\label{eq:q2b}
\bar{Q} = \left( 
\bar{a}^E_y - \bar{a}_y 
+ \dfrac{\bar{u}^E_y - \bar{u}_y}{\tau} 
\right) \dfrac{m}{E_y} \, .
\end{equation}
In this equation, the collisional forces and gas velocity vanish even though being implicitly included.
The operator $\bar{\phi}$ denotes the average over many particles at fixed points in space.
The particle velocities and accelerations in the above equation are measured in two separate experiments;
one experiment with the electric field switched off ($u_y$ and $a_y$) and one experiment with the electric field switched on ($u^E_y$ and $a^E_y$).
For the mathematical details of deriving equation~(\ref{eq:q2b}), we refer to the original publication~\citep{Gro23b}.

\subsection{Experimental results}

\begin{figure}[b]
\centering
\subfloat[]{
\includegraphics[trim=0mm 0mm 0mm 0mm,clip=true,width=0.47\textwidth]{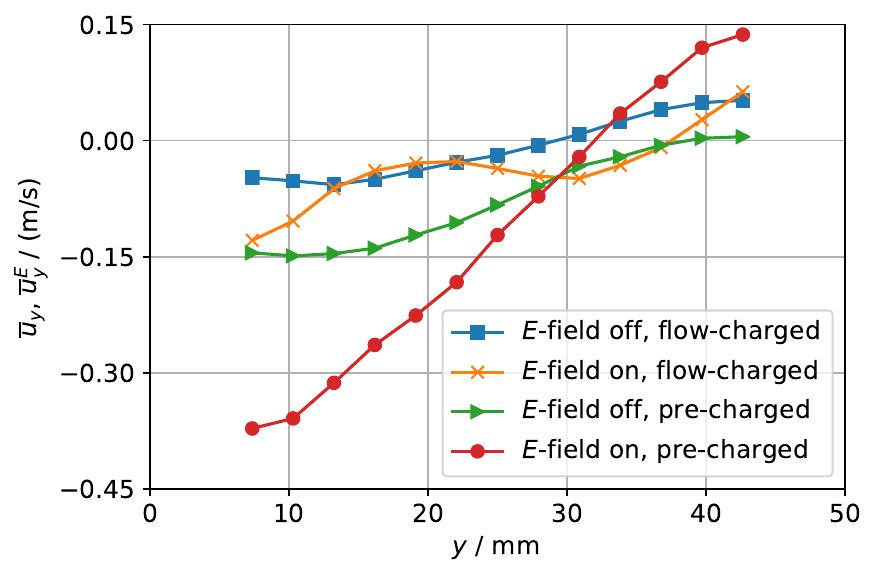}
\label{fig:u}
}
\subfloat[]{
\includegraphics[trim=0mm 0mm 0mm 0mm,clip=true,width=0.47\textwidth]{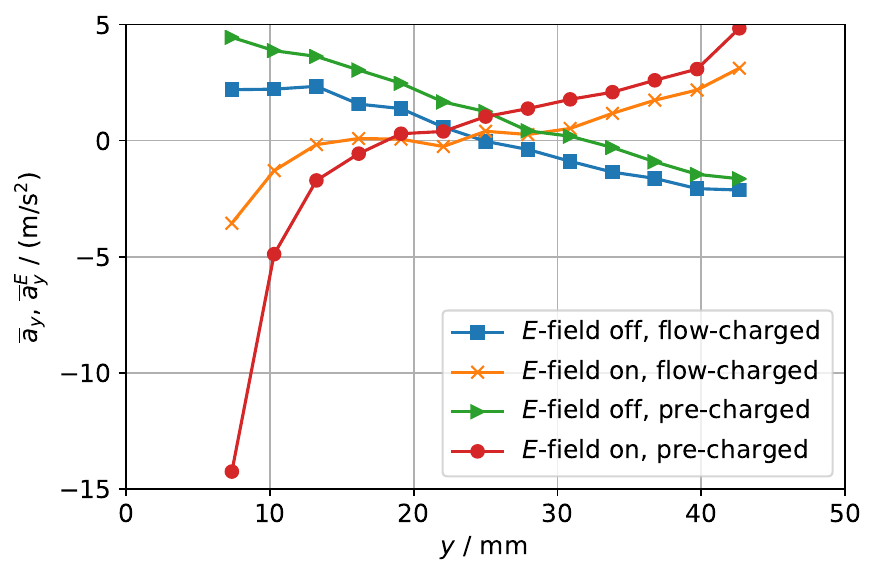}
\label{fig:a}
}
\caption[]{
Measured response of the particles to the electric field; 
Average particle wall-normal (a) velocities and (b) accelerations.
}
\label{fig:ua}
%\end{figure}
%\begin{figure}[b]
\begin{center}
\subfloat[]{\includegraphics[width=0.47\textwidth]{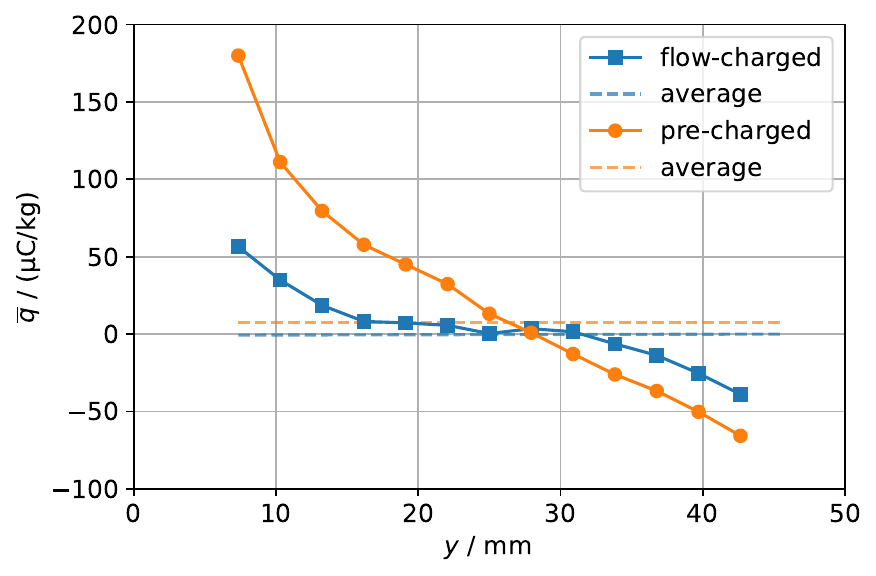}
\label{fig:qa}
}
\quad
\subfloat[]{\includegraphics[width=0.47\textwidth]{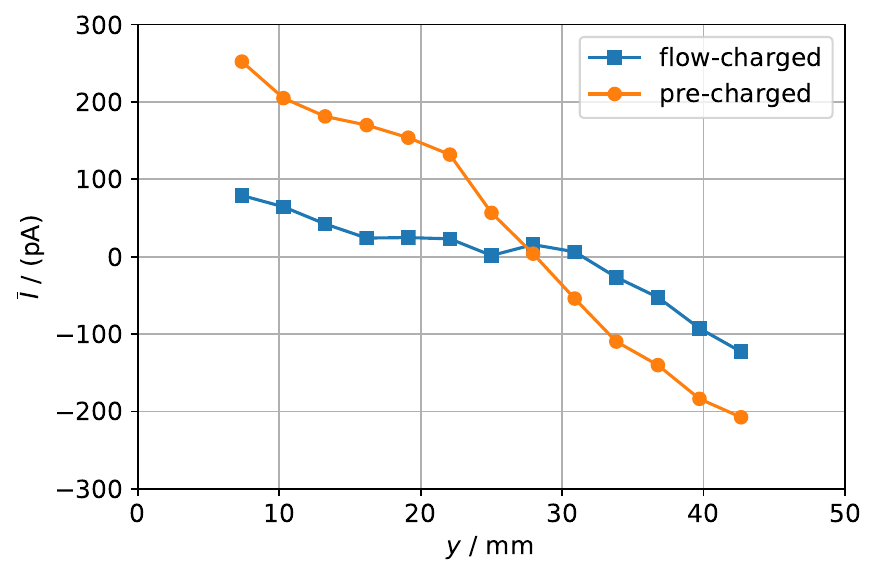}
\label{fig:qb}
}
\end{center}
\caption{Bipolar powder flow resolved with the new measurement system.
(a) Time-averaged specific particle charge compared to total average particle charge.
(b) Time-averaged electric current.
}
\label{fig:q}
\end{figure}

Figure~\ref{fig:ua} presents the particles' measured average velocities and accelerations.
Figure~\ref{fig:q} gives the particles' derived charge, in terms of the specific charge (\ref{fig:qa}) and the electric current (\ref{fig:qb}).

All data are time-averaged and depict a slice through the duct's centreline.
We succeeded resolving the flow for $7.4$\,mm $<y<42.6$\,mm.
Outside of this region, the particles frequently collide with the walls.
Within the electric field, the wall collision frequency changes, which violates the assumptions underlying equations~(\ref{eq:q2b}).
Further, the limited camera view close to the duct's walls led to the erroneous identification of particle trajectories.

To qualitatively validate the measurement system, we conducted two experiments, one with the pre-charge pipe and one without.
Without the pipe, the particles charge during feeding and flowing through the duct;
they are called \textit{flow-charged}.
With the pipe, the particles receive an additional pre-charge before entering the duct;
they are called \textit{pre-charged}.

%u,a
The average wall-normal velocities (figure~\ref{fig:u}) and accelerations (figure~\ref{fig:a}) fill the terms on the right-hand side of equation~(\ref{eq:q2b}).
Each data point averages 9000 to 80\,000 measured values. 

Downstream of the point-like feeding position, the particle flow widened up from the center toward the walls, as reproduced by the nearly symmetric velocity profile in figure~\ref{fig:u} of the flow-charged particles without the electric field.
The particle velocities and accelerations of the pre-charged particles responded stronger to the electric field than the flow-charged ones.

Based on the locally highest measured induced wall-normal acceleration, the flow-charged particles changed their location while passing the electric field in $y$ direction on the average up to 4.9\,mm and the pre-charged particles up to 5.6\,mm.
Thus, the measurement was invasive.
Moreover, this location change propagates to an uncertainty of the charge profiles' spatial coordinate in figure~\ref{fig:q}.
Reducing the downstream distance between the beginning of the electric field and the laser sheet would reduce spatial uncertainty.
However, the chosen distance ensures a low signal-to-noise ratio of the measured velocity and acceleration responses and the minuscule uncertainty of the derived charge, which we prioritized over spatial accuracy.

%q
The profiles in figure~\ref{fig:q} of both experiments reveal a bipolar charged particle flow.
For the flow-charged particles, the profile is nearly symmetric to the duct's centreline.
Close to the left wall, the particles carry a positive charge, and close to the right wall, they carry a negative charge.

The measured bipolar charge emanated from same-material contacts with the feeder and the duct's walls.
During feeding, the particles contacted the feeder's screw, which was already covered with other adhering particles.
Since also the duct is made of PMMA, the flow-charged particles underwent only same-material contacts.
Same-material contacts, lacking a net direction of charge transfer, resulted in the observed bipolar charging.

When passing the pre-charge pipe, the particles received, on average, a positive charge before entering the PMMA duct.
In the duct's left half, the pre-charged profile in figure~\ref{fig:q} qualitatively reflects the charge increase.

On the other hand, in the duct's right half, the pre-charged particles carried less average charge than the flow-charged particles.
Pre-charging did not simply offset the charge profile;
instead, being refined by the pipe and the assigned charge, changed the flow pattern.
An electric field affects the particles, even without the external field, for example, due to charge spots on the duct.
The positively charged particles moved toward the duct's left, resulting in the negative offset of the pre-charged particles' velocity profiles in figure~\ref{fig:u}.
When the positively charged particles moved to the left, but the negative ones remained, the local ($y>30$\,mm) average negative charge increased.
Thus, the pre-charged profile shows the charge increase and its influence on the particle concentration and the local charge distribution.

%3.39886029398614e-14 -3.9991370359584543e-16 1.083337124086556e-13 4.413837820966769e-15 -84.98984314428603 24.54410805355035
Additionally, figure~\ref{fig:q} displays the total average charge as a Faraday pail would detect it.
That means the average of the charge profiles weighted by the local particle concentration.
Pre-charging increased the average charge from -0.4~fC to 4.4 fC.
However, the resolved peak of the flow-charged particles is 85 times higher than the average, and the peak of the pre-charged particles is 25 times higher than the average.
Thus, a Faraday pail would dramatically underestimate the flow's charge.

\section{Conclusions}
Our simulations demonstrate that whichever powder type or charging mechanism, powder flows charge non-uniformly.
Non-uniform charge means some flow regions or particle size classes charge stronger or with a different polarity than others.
Conventional measurement equipment, such as Faraday pails and induction probes, give only the integral flow charge but fail to detect local charge peaks of both polarities.
Our newly developed measurement system can resolve the powder charge in space.
Thus, the new measurement system can detect local extreme charge peaks that were thus far hidden from measurements.
Even though these charge peaks arise during pneumatic conveying, they might propagate to downstream processes, storages, and silos and ignite the dust.
These hidden charge peaks might enlighten some of the large number of unexplained industrial dust explosions.
In any case, finding these peaks demonstrates that future research in explosion protection should not only focus on mitigating known hazards but, also, on revealing the hidden ones.

\section*{Acknowledgements}
This project received funding from the European Research Council~(ERC) under the European Union’s Horizon 2020 research and innovation programme~(grant agreement No.~947606 PowFEct).

{\setlength{\bibsep}{1pt}
\bibliography{\string~/essentials/publications}}

\end{document}